\documentclass{iopjournal}
\usepackage[utf8]{inputenc}
\usepackage{amsmath,amsthm,amssymb}
\usepackage{float}
\usepackage{graphicx}
\usepackage{dcolumn}
\usepackage{bm}
\usepackage{color}
\usepackage{hyperref}   
\usepackage[all]{hypcap}
\pdfoutput=1
\DeclareUnicodeCharacter{2212}{-}

\begin{document}
\articletype{Paper} 

\title{Propagation and collisionless damping of topologically-protected surface plasma waves in non-uniformly magnetized plasma columns}

\author{Roopendra Singh Rajawat$^{1,*}$ and  Gennady Shvets$^1$}

\affil{$^1$School of Applied and Engineering Physics, Cornell University, Ithaca, NY 14850}

\affil{$^*$Author to whom any correspondence should be addressed.}

\email{rupn999@gmail.com}


\begin{abstract}
Recent theoretical studies revealed the existence of topologically-protected surface plasma waves (TSPWs) in cold magnetized plasmas assumed uniform along the direction of a uniaxial magnetic field. Reflections-free propagation of the TSPWs along arbitrarily-shaped plasma boundaries oriented perpendicularly to the magnetic field was shown to be preserved even when their collisionless damping by localized upper-hybrid resonances was accounted for. Here we extend this theory to the realistic case of three-dimensional magnetic field produced by finite-sized magnetic coils. We demonstrate that when TSPWs are launched in the direction of the decreasing magnetic field, they are collisionlessly absorbed within a highly localized (evanescent) region as they propagate away from the magnetic coil. We show that the resulting wave reflection can be negligible -- in clear contrast with conventional wave reflection from the corresponding evanescence regions.
\end{abstract}

\section{Introduction}
A wide range of natural solid, gaseous, fluid, and plasma materials and metamaterials \cite{Hasan_rmp_2010,Goldman_np_2016,delplace_science_2017,Yang_prl_2015,
Parker_prl_2020b,fu_nc_2021} exhibit non-trivial topological properties that fundamentally impact wave propagation at the interfaces between topologically-distinct bulk domains. Specifically, the bulk-edge correspondence (BEC) principle \cite{Klitzing_prl_1980,Hatsugai_prl93,silveirinha_prb16,Gangaraj_prl_2020} predicts the existence and number of gapless unidirectional edge states that are spectrally co-located within a common bandgap of the two topologically distinct bulk materials, each of which is characterized by a different integer Chern number~\cite{Ozawa_rmp_2019} assigned to every propagation band. Non-trivial band topology (i.e. non-vanishing Chern number) requires the bulk materials to lack the time-reversal (TR) symmetry that can be broken by, for example, magnetic field or material motion. The gapless property of the edge states implies that they span the entire shared bulk bandgap while maintaining non-vanishing propagation speed throughout the bandgap. Therefore, such edge states are referred to as topologically robust: their propagation is immune to backscattering because of the absence of the edge states propagating in the opposite direction.

The existence and propagation properties of topological surface plasma waves (TSPW's) have been theoretically studied for gaseous plasmas under a set of simplifying assumptions~\cite{Parker_prl_2020b,Parker_jpp_2021,fu_nc_2021, Fu_jpp_2022}. Specifically, transversely bounded cold plasmas are assumed to be immersed in a static unidirectional axial magnetic field, and to be spatially uniform along the axial direction.  Even within the idealized assumptions, it has been found that, in a dramatic break with BEC, the TSPW's do not span across the entire topological bulk bandgap\cite{Parker_prl_2020b,fu_nc_2021,Gangaraj_prl_2020} -- a phenomenon referred to as a "BEC anomaly"~\cite{Gangaraj_prl_2020}. Our recent work~\cite{Rajawat_prl_2025} pointed out two key effects responsible for the BEC breakdown. One stems from the simplified assumption of an infinitely sharp domain wall between the plasma and vacuum regions\cite{Gangaraj_prl_2020}, which we refer to as hard plasma interface (HPI)\cite{Rajawat_prl_2025}, resulting in a quasi-electrostatic surface wave with an infinitely short wavelength along the domain wall. The frequency of such a mode can restrict the spectral region inside the bulk band accessible to the TSPW, thus violating the BEC\cite{Gangaraj_prl_2020,Rajawat_prl_2025}. While this unphysical effect can be suppressed by assuming a more realistic smooth plasma interface (SPI)~\cite{Parker_prl_2020b}, the spatial variation of the plasma density introduces additional physics such as a spatially varying upper-hybrid frequency and the emergence of an additional continuous spectrum of localized modes\cite{case_PoF60,briggs_PoF70,sedlacek_1971,Shvets_pop_1999}. As the result of this additional physics, proper undamped TSPWs were found only below the electron cyclotron frequency $\omega_{\rm c} =eB_0/m_{e} c$ of an electron rotating in a uniform magnetic field ${\bf B_0} =  B_0 \hat{z}$ (with $m_{e}/e$ equal to the electron mass/charge). Therefore, whenever $\omega_{\rm c}$ was located inside the topological bulk bandgap, TSPWs would not cross the entire bulk bandgap~\cite{Parker_prl_2020b,fu_nc_2021,Fu_jpp_2022}. This apparent violation of the BEC was resolved by our recent work that demonstrated the existence of collisionlessly-damped quasi-TSPWs \cite{Rajawat_prl_2025} that are nevertheless robust with respect to backscattering. Together with undamped TSPW's, these modes collectively span the whole band-gap, thereby restoring the BEC principle. Moreover, we have demonstrated that these modes are robust against reflections even in the presence of unidirectional sheared magnetic fields\cite{Rajawat_prl_2025}.

\begin{figure}[t]
    \centering
 \includegraphics[width=1\textwidth]{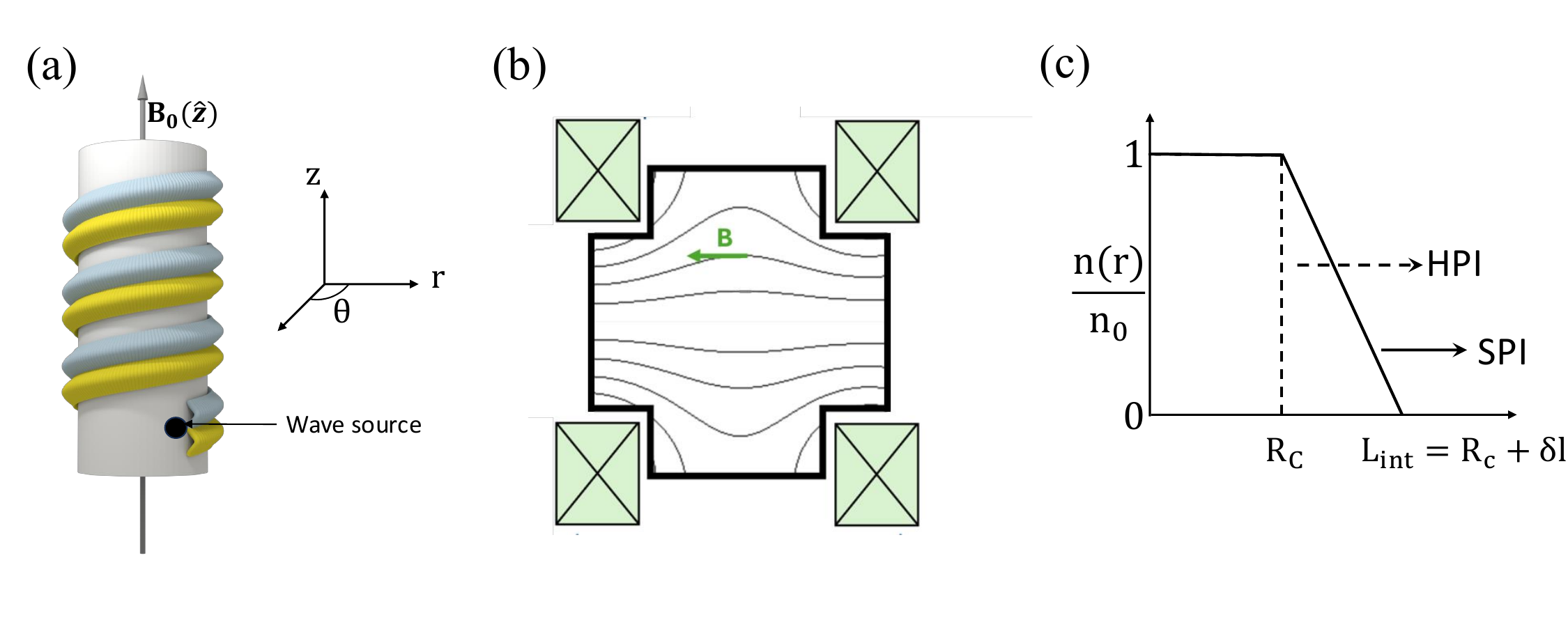}
        \caption{\label{fig:Fig1} (a) An illustration of a surface wave excitation by a point source (black disk) oscillating with frequency $\omega_{\rm dr}$, and its spiralling propagation at the interface between a magnetized plasma cylinder (gray) with radius $R_c$ and vacuum. (b) Field lines of a realistic non-uniform 3D magnetic field in a typical Penning discharge machine. (c) Radial plasma density profiles for hard (dashed line: HPI) and smooth (solid line: SPI) plasma interfaces.}
\end{figure}

In this work, we extend the earlier developed theories to account for the natural three-dimensional nature of the magnetic field created by realistic finite-sized current-carrying coils. The resulting magnetic field is neither unidirectional nor uniform. Even rotationally-symmetric around the axial $z$-axis finite-sized coils introduce a finite radial magnetic field component $B_{0r}(r,z)$ and the spatially-varying axial component $B_{0z}(r,z)$. An illustrative example of such a magnetic field configuration is shown in Fig. \ref{fig:Fig1}(b), which shows field distribution in a typical Penning discharge configuration used for confining neutral electron beam-produced plasmas~\cite{sakawa_pofb_1993,Chopra_apl_2025,Raitses_private_communication}. Naturally, the magnetic field generated by an L$2$ coil exhibits significant inhomogeneity in both the axial and radial directions. Therefore, it is important to understand the propagation properties of the "spiraling" (i.e. simultaneously propagating in the azimuthal and axial directions, as shown in Fig.~\ref{fig:Fig1}(a)) TSPW's when the translational symmetry in the axial $z$-direction is broken, e.g., through the non-uniformity of a realistic magnetic field with expanding magnetic field lines as shown in Fig.\ref{fig:Fig1}(b).

The rest of the manuscript is organized as follows. In Section~\ref{section:theoretical model}, we develop semi-analytical model for collisionlessly damped propagation of TSPWs around a uniformly magnetized cylindrical plasma columns separated from the surrounding vacuum region by a smooth density transition. This calculation extends our previously developed theory for slab-shaped (i.e., infinitely extended in a plane parallel to the uniaxial magnetic field ${\bf B_0} =  B_0 \hat{z}$) to azimuthally symmetric (round, with a finite radius) plasma cylinders that are translationally invariant in the direction of $\hat{z}$ of the magnetic field. Using semi-analytic calculations and particle-in-cell (PIC) simulations, we demonstrate that for the frequencies $\omega$ above the cyclotron frequency $\omega_{\rm c}$, such waves undergo spatial decay in both $z$-  $\theta$-directions as they spiral away from their localized monochromatic source. The propagation of TSPW's in a realistic non-uniform magnetic field of a finite-sized current-carrying coil is considered in Section~\ref{section:expandingB}. We show that when a TSPW propagate in the axial direction of a decreasing axial magnetic field, it undergoes complete collisionless absorption in those regions of the plasma column where the local electron cyclotron frequency $\omega_{\rm c}(z)$ falls below the frequency $\omega$ of the launched TSPW. We further demonstrate that such complete (i.e. reflection-free) absorption is possible even when the spatial scale of the magnetic field decrease is shorter that the radius of the plasma column. Our findings are summarized in Section \ref{section:conclusions}.

\section{Theoretical Model} \label{section:theoretical model}
As a model for investigating topological surface plasma waves, we use a physical configuration shown in Fig.~\ref{fig:Fig1}(b): a cylindrical magnetized plasma with density $n(\mathbf{r}) \equiv n(r)$ (where $n(r) \rightarrow n_0$ for $r \ll L_{\rm int}$) interfacing with a vacuum region at $r \approx L_{\rm int}$. Therefore, the density of the plasma column is assumed to be uniform and equal to $n_0$ inside its effective radius $R_c$, and to uniformly fall down to zero over a distance $\delta l \ll R_c$.  Next, we derive the TSPW dispersion relation in the form of $\omega \equiv \omega(m,k_z)$, where $\omega$ is a complex-valued (i.e., possibly decaying) frequency of the wave, while $m$ and $k_z$ are the integer azimuthal and real-valued axial wave numbers, respectively. To derive the dispersion relation, we first obtain the solutions to Maxwell’s equations in the bulk of the plasma column ($r < R_c$) and in the surrounding vacuum ($r > R_c + \delta l$). These solutions are then matched by integrating Maxwell’s equations across the narrow finite-width plasma–vacuum interface of the thickness $\delta l$, yielding the desired dispersion relation. For the HPI ($\delta l = 0$), the continuous field components of the TSPW are matched at $r=R_c$ to obtain the corresponding real-valued dispersion relation for $\omega \equiv \omega^{\rm HPI}(m,k_z)$.
\subsection{Formulation of the propagation equations for TSPWs in spatial/temporal frequency domains}
We assume that TSPW's electric ($\bf E$) and magnetic ($\bf B$) fields are harmonic in time and space, i.e. proportional to $e^{i k_z z - i \omega t}$. After space-time Fourier transformation of Maxwell equations \cite{Jackson}, we obtain:
\begin{align}
{\bf \nabla} \times {\bf E} & =  i k_0 {\bf B}, \label{eq1}\\
 {\bf \nabla } \times {\bf B} & =  -i k_0 \hat{\epsilon} {\bf E}, \label{eq2}
\end{align}
where the dielectric tensor $\hat{\epsilon}$ for a cold magnetized plasma is given by \cite{Stix}
\begin{equation*}
  \epsilon =
\begin{pmatrix}
\epsilon_t & - \epsilon_{g} & 0  \\
 \epsilon_{g} & \epsilon_{t} & 0 \\
0 & 0 & \epsilon_{a}
\end{pmatrix}.
\end{equation*}

Here $\omega_{p0} = \sqrt{4 \pi e^2 n_{0}/m_{e}}$ is the peak electron plasma frequency inside the column, $c$ is the speed of light, $ \epsilon_{t}  =  1 - \frac{\omega^{2}_{p0} }{ \omega^{2} - \omega^{2}_{c}},
    \epsilon_{g}  = i \frac{\omega_{c}}{\omega}  \frac{\omega^{2}_{p0}}{\omega^{2}_{c} - \omega^{2}},
    \epsilon_{a}  =  1 - \frac{\omega^{2}_{p0}}{\omega^2 } $, and $k_0 = \omega/c$. The remaining electromagnetic field components ($E_r, E_\theta, B_r, B_\theta$) of the TSPW can be expressed in terms of its axial fields and their radial derivatives through the following matrix \cite{Margot_jpd_1991}:
\begin{eqnarray}
\begin{pmatrix}
E_r \\  E_\theta \\ B_r \\ B_\theta
\end{pmatrix}
= D^{-1} \begin{pmatrix}
A_{11} & -A_{12} & A_{31} & A_{32} \\
A_{12} &  A_{11}  & -A_{31} & A_{31} \\
A_{13} & -A_{14} & -A_{11} & A_{12} \\
A_{14} &  A_{13}  &- A_{12} & -A_{11} \\
\end{pmatrix}
\begin{pmatrix}
- i m E_z/r \\  \partial E_z/\partial r \\ i m r  B_z/r \\ - \partial B_z/\partial r
\end{pmatrix}
\end{eqnarray}
where the matrix elements $A_{ij}$ can be expressed in terms of the $r$-dependent components of the dielectric tensor of the cold magnetized plasma~\cite{Stix}: $ \epsilon_{t} \left(r,\omega \right)  =  \frac{\omega^2 - \omega^{2}_{\rm UH}(r)}{ \omega^{2} - \omega^{2}_{c}}$, $\epsilon_{g} \left(r, \omega \right) =
 i \frac{\omega_{c}}{\omega}  \frac{\omega^{2}_{p}(r)}{\omega^{2}_{c} -\omega^{2}}$, and $\epsilon_{a} \left(r \right) = \frac{\omega^2 - \omega^{2}_{p}(r)}{\omega^2}$.
Specifically, $A_{11} = i k_0^2 \epsilon_g k_z$,
$A_{12}  =   i k_z \left( k_0^2 \epsilon_a - k_z^2  \right)$,
$A_{13} = -i \omega \left[ k_0^2 \left( \epsilon_t^2 - \epsilon_g^2 \right) - \epsilon_t k_z^2 \right]$,
$A_{14} =  \omega k_z^2  \epsilon_g$,
$A_{31} =  -i \omega \left( k_0^2 \epsilon_t - k_z^2  \right)$,
$A_{32} =  - \omega k_0^2 \epsilon_g$, and $D = (k_0^2 \epsilon_t - k_z^2)^2 - k_0^4 \epsilon_g^2$.

Equations \eqref{eq1} and \eqref{eq2} can be further simplified to \cite{Allis_1963,Margot_jpd_1991}
\begin{eqnarray}
\nabla_T^2 E_z - a E_z  = b B_z, \label{eq3}  \\
\nabla_T^2 B_z - c B_z = d E_z, \label{eq4}
\end{eqnarray}
where $\nabla^2_{T} \equiv \partial^2/\partial r^2 + r^{-1} \partial/\partial r + r^{-2}\partial^2/ \partial \theta^2$ is the transverse Laplacian in cylindrical coordinates $(r,\theta)$, $a = (k_z^2 - k_0^2 \epsilon_t ) \epsilon_a/\epsilon_t $, $b = - k_0 k_z \epsilon_g/\epsilon_t$, $c = k_z^2 - k_0^2 \epsilon_{tg}^2 / \epsilon_t$ and $d = k_0 k_z \epsilon_a \epsilon_g / \epsilon_t$. After lengthy but straightforward algebraic manipulations, we obtain a set of de-coupled forth order differential equations for the axial electric and magnetic fields of the TSPW:
\begin{eqnarray}
\left[ (\nabla_T^4 - (a+c) \nabla_T^2 + (ac-bd) \right] E_z  = 0,  \label{eq5}  \\
\left[ (\nabla_T^4 - (a+c) \nabla_T^2 + (ac-bd) \right] B_z  = 0. \label{eq6}
\end{eqnarray}
By introducing auxiliary constants $p_1$ and $p_2$ as
\begin{eqnarray}
p_1^2 + p_1^2 = a + c, \label{eq7}  \\
p_1^2 p_2^2 = ac-bd,	\label{eq8}
\end{eqnarray}
the equations are further de-coupled to yield
\begin{equation} \label{eq9}
\left( \nabla_T^2 - p_1^2 \right) \left( \nabla_T^2 - p_2^2 \right) \left( E_z, H_z \right) = 0
\end{equation}
for the axial field components of the TSPW. The solution of Eq.~(\ref{eq9}) for the field components proportional to $e^{i m \theta}$ (where $ \rm m$ is the azimuthal wavenumber) satisfying the regularity condition at $r = 0$ can be written as
\begin{equation} \label{eq10}
E_z = \left[k^{\rm (p)}_1 I_m \left(p_1 r \right) + k^{\rm (p)}_2 I_m \left(p_2 r \right) \right]e^{i m \theta},
\end{equation}
where \( k^{\rm (p)}_{1,2} \) are arbitrary coefficients in the bulk plasma, $I_m$ is the modified Bessel function of first kind of $\mathrm{m^{th}}$ order and $p_{\lbrace i = 1,2 \rbrace}$ are given by
\begin{equation} \label{eq11}
2 p^2_{1,2} = \left( a + c \right) \pm \left[(a + c)^2 - 4(ac-bd)  \right]^{1/2}.
\end{equation}
Similarly, the solution for $B_z$ can be written as
\begin{equation} \label{eq12}
B_z = \left[k^{\rm (p)}_1 h_1 I_m \left(p_1 r \right) + k^{\rm (p)}_2 h_2 I_m \left(p_2 r \right) \right]e^{i m \theta},
\end{equation}
where $h_{\lbrace i = 1,2 \rbrace} = a+p_{\lbrace i = 1,2 \rbrace}/b$.

In vacuum, equations \eqref{eq3} and \eqref{eq4} are de-coupled and simplified to
\begin{eqnarray}
\left(\nabla_T^2 - \kappa_0^2 \right) E_z  = 0 \label{eq13}   \\
\left(\nabla_T^2 - \kappa_0^2 \right) B_z = 0  \label{eq14}
\end{eqnarray}
where $\kappa_0^2 = k_z^2 - k_0^2$ and solutions for $E_z$, $B_z$ can be written as
\begin{eqnarray}
E_z = k^{\rm (v)}_1 K_m \left(\kappa_0 r \right)e^{i m \theta}, \label{eq15}  \\
B_z = k^{\rm (v)}_2 K_m \left(\kappa_0 r \right)e^{i m \theta}. \label{eq16}
\end{eqnarray}
where \( k^{\rm (v)}_{1,2} \) are arbitrary coefficients in vacuum and  $K_m$ is the modified Bessel function of second kind of $\mathrm{m^{th}}$ order.

\subsection{Derivation of the dispersion relations for TSPWs from the boundary conditions}

From Eqs.(\ref{eq10},\ref{eq12},\ref{eq15},\ref{eq16}), we observe that the spatial profile of every TSPW is characterized by a set of $4$ numbers: $\left( k^{\rm (v)}_{1,2} \right)$ and $ \left( k^{\rm (p)}_{1,2} \right)$. Because of the linear nature of the Maxwell's equations, one of these coefficients can be arbitrarily chosen. By matching the bulk plasma and vacuum (super-scripts ${\rm (p)}$ and ${\rm (v)}$, respectively) across the plasma-vacuum interface, the dispersion relations for the TSPWs supported by either a hard plasma-vacuum interface (HPI-TSPW) or a smooth plasma-vacuum interface (SPI-TSPW): see Fig.~\ref{fig:Fig1}(c).

For the HPI, an analytic expression in the form of the dispersion function $D^{\rm HPI}(\omega,m,k_z) = 0$ can be obtained and used to calculate the dispersion relation $\omega^{\rm HPI}(m,k_z)$. To calculate $D^{\rm HPI}$, we match the analytically known bulk plasma and vacuum field profiles at $r=R_c$. This is accomplished by introducing a four-component vector $\mathbf{\psi} \equiv \left(B_r,B_{\theta},B_z,E_z\right)^T$ comprising the field components remaining continuous across the discontinuous HPI. Using the corresponding bulk plasma and vacuum vector states $\mathbf{\psi}^{\rm (p,v)}_{1,2}$, the dispersion function can be expressed as the generalized Wronskian of the $\mathbf{\psi}$-vectors evaluated at $r=R_c$\cite{Rajawat_prl_2025}:

\begin{equation}\label{eq:wronskian}
  D^{\rm HPI}(\omega,m,k_z) = \det \left(\mathbf{\psi}^{\rm (p)}_{1} \mathbf{\psi}^{\rm (p)}_{2} \mathbf{\psi}^{\rm (v)}_{1} \mathbf{\psi}^{\rm (v)}_{2} \right) \left|_{r=R_c} \right.
\end{equation}

\begin{figure}[t!]
    \centering
    \includegraphics[width=1\textwidth]{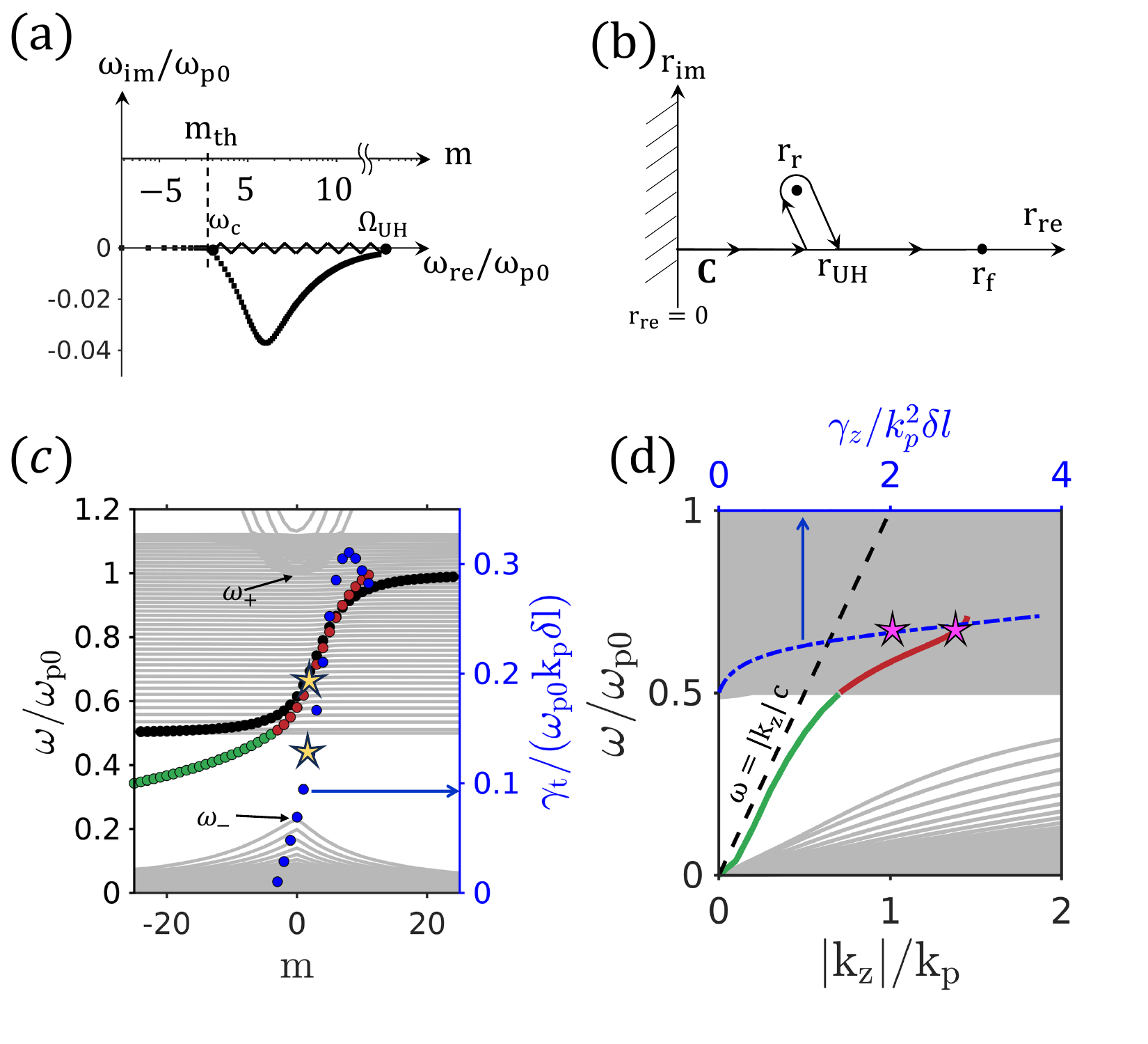}
    \caption{(a,b) Integration contours in the complex (a) $\omega$- and  (b) $r$- planes used for calculating the complex-valued dispersion relation $\omega - i \gamma_t \equiv \omega^{\rm SPI}\left( m,k_z={\rm const} \right)$. (c,d) Dispersion relation curves for the modes supported by a magnetized plasma cylinder with radius $R_c$. Common plasma parameters: $\omega_c = 0.5 \omega_{p0}$, $\mathrm{R_c = 5 k_p^{-1}}$, and $\delta l = 0.2 k_p^{-1}$ (for SPI). Gray curves: propagating bulk and ramp-localized continuum modes. (c)Complex-valued frequency $\tilde{\omega} \left(m,k_z=k_p \right) \equiv \omega - i\gamma_t$ plotted as a function of the variable azimuthal number $m$.   Undamped surface waves: HPI-TSPW (solid black circles)and SPI-TSPW (solid green circles). Damped SPI-TSPW quasi-modes: temporal oscillation frequency ($\omega$: solid red circles) and decay constant ($\gamma_t$:  solid blue circles). Yellow stars correspond to the $(m=1,k_z=k_p)$ TSPW. Lower (upper) bounds of the bulk bandgap are marked as $\omega_{-}$ ($\omega_{+}$). (d) Complex-valued axial wavenumber $\tilde{k}_z^{\rm SPI} \left(m=0,\omega \right) \equiv k_z + i\gamma_z$ plotted as a function of the real-valued wave frequency $\omega$. Undamped SPI-TSPW (green line) and damped SPI-TSPWs: propagation ($k_z$: red line) and spatial decay ($\gamma_z$: blue line) constants. Pink stars correspond to $\omega = 0.67\omega_{p0}$.}
    \label{fig:Fig2}
\end{figure}

For an HPI-TSPW, the resulting dispersion relation curve $\omega \equiv \omega^{\rm HPI}(m,k_z)$ is plotted with filled-black circles in Fig.~\ref{fig:Fig2}(c). Note that the dispersion curve does not span the entire bandgap: it flattens at $m \rightarrow \pm \infty$. This unphysical flattening is a result of our idealized assumption of the infinitely-sharp plasma-vacuum interface: it disappears for the realistic SPIs~\cite{Rajawat_prl_2025}.

The dispersion function $D^{\rm SPI}(\omega,m,k_z) = 0$ for the smooth plasma interface ($\delta l \neq 0$) is obtained in a similar way, except that corresponding vector states $\mathbf{\psi}^{\rm (p,v)}_{1,2}$ are evaluated at $r=L_{\rm int}$. Therefore, the functional form of the vacuum states $\mathbf{\psi}^{\rm (v)}_{1,2}$ is unchanged while the plasma states $\mathbf{\psi}^{\rm (p)}_{1,2}$ must be propagated from their plasma bulk values at $r=R_c$ derived from Eqs.(\ref{eq10},\ref{eq12}) to their plasma edge values at $r=L_{\rm int}$. Such propagation of the $\mathbf{\psi}$-states across the SPI layer is done by integrating the following differential equation\cite{Rajawat_prl_2025}:

\begin{equation}\label{eq:shooting_method}
  \frac{\partial \mathbf{\psi}^{\rm p}}{\partial r}   = \frac{i}{ r \epsilon_t(r,\omega)} M \mathbf{\psi}^{\rm p},
\end{equation}
where the $4 \times 4$ matrix $M\left(r,\mathbf{m},\omega \right)$ is given by
\begin{equation} \label{eq17}
M = \begin{pmatrix}
i \epsilon_t & - m \epsilon_t & - r k_z \epsilon_t & 0 \\
 m \epsilon_t  & i \epsilon_t & 0 & - r k_0 \epsilon_a \epsilon_t \\
 r k_z  \epsilon_t -  r \frac{k_0^2}{k_z}  \epsilon_{tg}^2  &  r k_z \epsilon_g & - m \epsilon_g &  m \frac{k_0}{k_z}  \epsilon^2_{tg} \\
- r k_0  \epsilon_g & r \frac{ k_z^2 - k_0^2 \epsilon_t  }{k_0} & - m \frac{k_z}{ k_0}   &  m \epsilon_g
\end{pmatrix}.
\end{equation}
The expressions for the $r$- and $\omega$-dependent diagonal ($ \epsilon_{t}$ and $\epsilon_{a}$) and off-diagonal ($\epsilon_{g}$) components of the dielectric tensor of the cold magnetized plasma were listed above, and  $\epsilon^2_{tg} \equiv \epsilon^2_t + \epsilon_g^2$.

By integrating Eq.(\ref{eq:shooting_method}) between $r=R_c$ and $r=L_{\rm int}$ endpoints and substituting thus obtained values of $\mathbf{\psi}^{\rm (p)}_{1,2}(r=L_{\rm int})$ into Eq.(\ref{eq:wronskian}) results in a dispersion function $D^{\rm SPI}(\omega,m,k_z)$. The corresponding dispersion relation $\omega = \omega^{\rm SPI}(m,k_z)$ is implicitly contained in the $D^{\rm SPI}(\omega,m,k_z) = 0$ condition that follows from the continuity of the plasma and vacuum fields at the $r=L_{\rm int}$ plasma/vacuum interface.

When the eigen-frequencies of the SPI-TSPW modes fall inside the $\omega_{-} < \omega < \omega_{c}$ frequency range (where $\omega_{-}$/$\omega_{+}$ are the lower/upper edges of the bulk bandgap, see Fig.~\ref{fig:Fig2}(c)), the $r$-integration of Eq.(\ref{eq:shooting_method}) can be carried out along the real axis because no zeros of $\epsilon_t(r,\omega)$ are encountered for any $R_c < r < L_{\rm int}$ for any plasma density profile $n_0(r)$. The real-valued (damping-free) dispersion relation curve $\omega(m,k_z={\rm const})$ \cite{Parker_prl_2020b,fu_nc_2021} is plotted with green filled-circles in Fig.~\ref{fig:Fig2}(c) (where $k_z=k_p \equiv \omega_{p0}/c$) as a discrete function of the integer azimuthal wavenumber $m$. For the plasma parameters listed in the caption of Fig.~\ref{fig:Fig2}, such real-valued eigen-mode frequency solutions exist only for $m \leq m_{\rm th} = -4$.

The situation changes dramatically for $m > m_{\rm th}$ because the frequency $\omega$ of the TSPW mode moves above the cyclotron frequency, and is no longer real because of the phenomenon of continuum damping~\cite{briggs_PoF70,sedlacek_1971,Chu_pop_1994,Shvets_pop_1999} ubiquitous in plasmas. Continuum damping manifest itself in collisionlessly-damped quasimodes (QM) generated by
phase mixing of multiple local upper-hybrid resonances~\cite{Rajawat_prl_2025}. Formally, the QMs emerge because of the  $1/\epsilon_t(r,\omega)$ factor in Eq.(\ref{eq:shooting_method}) acquiring a pole singularity at the local upper-hybrid resonance location $r=r_{\rm UH}(\omega)$ defined by $\omega^2 = \omega^2_{c} + \omega^2_p (r_{\rm UH})$ as soon as the $\omega > \omega_c$ condition is satisfied. The ambiguity of integrating Eq.(\ref{eq:shooting_method}) across the singularity is resolved by analytically continuing the dispersion function ($D \rightarrow D_{\ast}$) into the lower-half of the complex $\omega$-plane to preserve causality~\cite{briggs_PoF70,strogatz_prl92,Shvets_pop_1999}. If $D_{\ast}$ possesses a pole $\tilde{\omega} \equiv \omega -i \gamma_{\rm t}$ (where $\omega$ and $\gamma_{\rm t}$ are both real-valued, and $\gamma_{\rm t} > 0$), then its analyticity requires that the integration between $r = R_c$ and $r=L_{\rm int}$ be carried out along the path $\mathcal{C}$ in the complex-$r$ plane passing above the complex-valued point $r_{\rm r}$ defined as $\omega_{\rm UH}(r_{\rm r}) = \tilde{\omega}$~\cite{briggs_PoF70,Shvets_pop_1999}: see Fig.~\ref{fig:Fig2}(b).

The complex-valued roots $\tilde{\omega}$ of the dispersion function $D_{\ast}(\tilde{\omega},m,k_z)=0$ yields the dispersion relation $\tilde{\omega}(m,k_z)$ for any fixed real-valued $k_z$ and any SPI density profile $n_0(r)$. In Fig.~\ref{fig:Fig2}(c), we present examples of the complex frequency $\tilde{\omega}(m) \equiv \omega - i\gamma_t$ assuming the SPI in the form of a linearly-varying plasma density ramp extending from $r_{\rm in}=R_c$ to $L_{\rm int} = R_c + \delta l $: $n(r) = n_0(1 -\left( r -R_c \right) / \delta l)$ and $\delta l = 0.2 k_p^{-1}$. Note that the resonant point $r_{\rm re} \equiv r_{\rm UH}$ (where  $r_{\rm UH}/\delta l = \left( \omega_{\rm UH}^2 - \omega^2 \right)/\omega_{p0}^2$) is also real-valued, so the integration contour $\mathcal{C}$ includes a semi-circle above $r=r_{\rm UH}$. The numerically calculated oscillation (decay) constants $\omega$ ($\gamma_t$) are plotted in Fig.~\ref{fig:Fig2}(a) as filled-red (filled-blue) circles, respectively. While the $\omega(m)$ curves for the HPI- and SPI-TSPW overlap for small values of $m$, continuum damping removes unphysical flattening of HPI-TSPW dispersion curve for large azimuthal mode numbers, thereby "rescuing" the BEC principle.

Similarly, the numerically calculated spatial oscillation (decay) constants $k_z$ ($\gamma_z$) for axial wave propagation can be calculated for a fixed azimuthal number $\rm ( m)$ by finding zeros of the dispersion function $D_{\ast}(\omega,m,\tilde{k}_z)=0$ assuming that $\omega$ is real-valued and $\tilde{k}_z = k_z + i \gamma_z$. One such HPI-TSPW dispersion curve is plotted in Fig. \ref{fig:Fig2}(d) for an azimuthally symmetric ($\rm m =0$) mode. The rest of the plasma parameters listed in the caption of Fig.~\ref{fig:Fig2} are common with Fig.\ref{fig:Fig2}(c). Note that because $m > m_{\rm th}$, the quasi-mode is weakly-evanescent: its small spatial and temporal evanescence coefficients $\gamma_z$ and $\gamma_t$ are related through the finite group velocity of the TSPW: $\gamma_t = \gamma_z v_g$, where $v_g = d\omega/dk_z \approx 0.3$. Finally, the dispersion curve $\omega(m=0,k_z)$ of the collisionlessly damped TSPW quasi-mode lies below the light line defined as $\omega^2 = k_z^2c^2$ and shown as a dashed black line in Fig.~\ref{fig:Fig2}(d). That implies that the damping of the quasi-mode is not caused by its leakage into the vacuum region. Instead, the collisionless damping occurs because of the mode coupling to the continuum of localized upper-hybrid resonances inside the density ramp~\cite{Rajawat_prl_2025}.

It should also be noted that the dispersion curves $\omega_m(k_z)$ plotted in Fig. \ref{fig:Fig2}(d) are symmetric in $ \pm k_z$. The existence of the same-frequency TSPWs with the wave numbers $\pm k_z$ propagating in the opposite directions has important implications for the TSPW propagation in axially nonuniform magnetized plasma. When the translational symmetry in the axial direction is broken because of the $z$-dependence of either the external magnetic field or the plasma density (or both), the axial wavenumber is no longer conserved. That implies the possibility of backscattering: from a forward-propagating TSPW with $k_{\rm forw} = k_z > 0$ into a backward-propagating TSPW with $k_{\rm back} = -k_z$. The coexistence of the surface waves that are backward- and forward-propagating in the axial direction stands in contrast with the existence of one-way TSPWs that can propagate only in one azimuthal direction. Moreover, azimuthal symmetry is not a precondition for unidirectional backscattering-free propagation of TSPWs around the plasma column: even non-round cylindrical plasma columns support such waves owing to the BEC. On the other hand, TSPWs retain their topological protection from backscattering only for the magnetized plasma columns that are translationally invariant in the axial direction.

\subsection{Particle in Cell (PIC) simulations of TSPWs}
\begin{figure}[t!]
    \centering
    \includegraphics[width=0.8\textwidth]{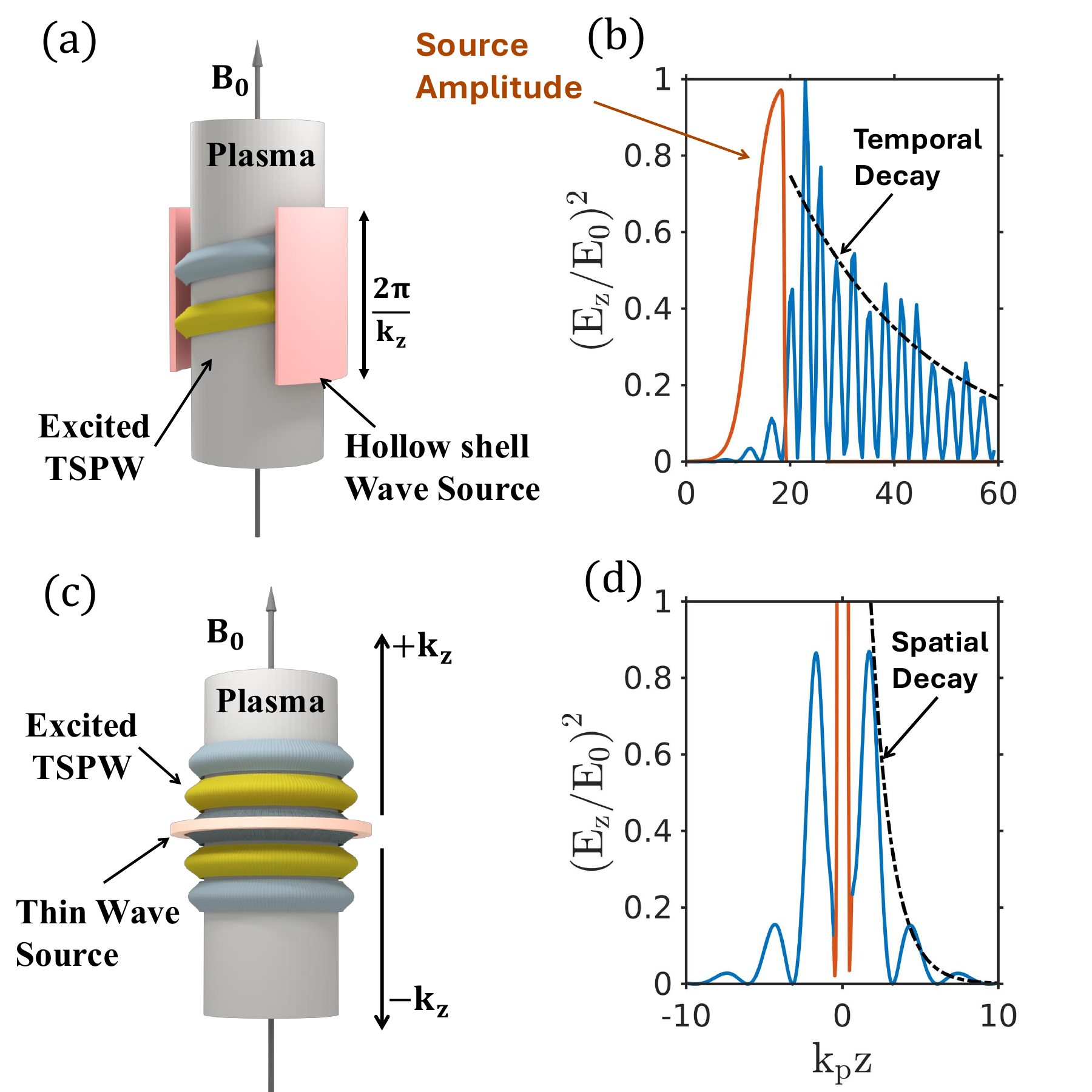}
    \caption{(a,c) An illustration of TSPW excitation in magnetized plasma columns with an SPI using a current-carrying shell that is either spatially periodic (a) or thin (c) in the axial $z$-direction. The applied current pulse is temporally short in (a) or periodic in (c), with the drive frequency $\omega_{\rm dr} = 0.62 \omega_{p0}$ . (b,d) Temporal (b) and spatial (d) decay of the electric field intensity $\mathrm{E_z^2}$ (solid blue line), and comparisons with the corresponding theoretical damping rates (dot-dashed black line) for the TSPW quasi-mode with $m=1$ (b) and $m=0$ (d). Red solid lines: temporal (b) and spatial (d) dependencies of the wave launching source amplitudes. The central frequencies of the source are (b) $\omega_{\rm  dr} = 0.62\omega_{p0}$ (yellow star in Fig. \ref{fig:Fig2}(c)) and (d) $\omega_{\rm  dr} = 0.67\omega_{p0}$ (pink star in Fig. \ref{fig:Fig2}(d)). Common plasma parameters: same as in Fig.~\ref{fig:Fig2}.}
    \label{fig:Fig22}
\end{figure}

To validate the semi-analytic theory developed earlier in the Section \ref{section:theoretical model}, we next use the first-principles three-dimensional particle-in-cell (3D-PIC) code VLPL~\cite{pukhov_1999} to confirm the existence of (i) temporally-damped TSPWs with a fixed axial wavenumber $k_z$ (see Figs.\ref{fig:Fig22}(a,b) for illustration), and (ii) spatially-evanescent TSPWs with a fixed frequency $\omega$ (see Figs.\ref{fig:Fig22}(c,d)). Each simulation setup ensures a fixed integer azimuthal wavenumber $m$ of the mode, as well as real-valued $k_z$ (in simulation (i): see Fig.\ref{fig:Fig22}(a) for a schematic) and $\omega$ (in simulation (ii): see Fig.\ref{fig:Fig22}(c)). For both computational demonstrations, we choose common plasma parameters listed in the caption of Fig.~\ref{fig:Fig2}(c,d). See appendix for additional simulations. 

\subsubsection{PIC simulation (i): temporal decay of a TSPW with fixed $k_z$ and $m=1$}

To demonstrate temporal damping, we choose simulation domain size $\mathrm{L_x = L_y \times L_z} =  \mathrm{20 ~\mathrm{k_p^{-1}} \times 2 \pi \, ~\mathrm{k_p^{-1}}}$, with absorbing boundary condition in radial direction and periodic boundary condition in \(z-\)direction. To selectively drive TSPWs with fixed \( \left( k_z=k_p, m=1 \right) \) and central frequency $\omega=\omega_{\rm dr}$, a cylindrical hollow shell wave source  surrounding the SPI is introduced, as indicated by a pink-colored shell in Fig.~\ref{fig:Fig22}(a). The axial length $L_z = 2 \pi /k_z$ is chosen to ensure translational symmetry and infinite spatial extent of the excited TSPW.

The time-dependent excitation is implemented using a surface current over a cylindrical shell and varying in space-time according to \[J_z(t,z,\theta) \propto 0.5 \left( \tanh \left(\frac{t-t_{1}}{t_{\rm rise}} \right) + \tanh \left( \frac{t_{2}-t}{t_{\rm fall}} \right) \right) \cos \left( k_z z + \theta - \omega_{\rm  dr} t \right) \]. In our simulation we have used the following parameters for the beginning, end, rise, and fall times of the excitation pulse: \(\omega_{p0} t_1 = 15, \) \( \omega_{p0} t_2 = 20, \) \( \omega_{p0} t_{\rm rise} = 3 \) and \( \omega_{p0} t_{\rm fall} = 0.2  \). The amplitude of the source envelope is plotted with a solid red line in Fig. \ref{fig:Fig2}(b).

Because the central frequency of the excited mode satisfies the $\omega_{\rm dr} > \omega_c$ condition, it is expected to undergo collisionless post-excitation damping as indicated by yellow stars in Fig.~\ref{fig:Fig2}(c). This theoretical prediction is indeed confirmed by the plot of $\rm E_z^2$ (blue line in Fig.~\ref{fig:Fig22}(b)) as a function of time at a fixed radial position. The temporal damping rate $\gamma_t \approx 0.019 \omega_{p0}$ obtained from our semi-analytic model (yellow star on filled blue circles in Fig.~\ref{fig:Fig2}(c)) shows excellent agreement with the temporal damping rate observed in the simulation (fitted black dot-dash line in Fig.~\ref{fig:Fig22}(b)).

\subsubsection{PIC simulation (ii): spatial evanescence in the axial direction of a monochromatic $m=0$ TSPW}
Next, we demonstrate the existence of spatially damped TSPW using a 3-D PIC simulation. This demonstration validates the results of the semi-analytic quasi-mode theory shown in Fig.~\ref{fig:Fig2}(d). The computational domain size $\mathrm{L_x = L_y = L_z} =  \mathrm{35 ~\mathrm{k_p^{-1}}}$ is chosen, with absorbing boundary conditions implemented for all six domain boundaries. The rest of the plasma bulk and ramp parameters are identical to those used in Fig.~\ref{fig:Fig2}. A monochromatic azimuthally symmetric current source for exciting the $m=0$ surface wave with $\omega = \omega_{\rm dr}$ is implemented on a thin annular ring indicated by a pink-colored ring in Fig. \ref{fig:Fig22}c). Both forward and backward $\pm z$-propagating TSPWs of equal amplitudes are launched by the source, provided that the ring is sufficiently thin in the axial direction and \(J_z(t) \propto \cos \left(\omega_{\rm  dr} t \right). \) The drive frequency $\omega_{\rm dr} = 0.67 \omega_{p0}$ corresponds to the pink stars in Fig.~\ref{fig:Fig2}(d).

The electric field intensity plotted in Fig.~\ref{fig:Fig22}(d) by a blue-colored line indeed confirms that both identical counter-propagating waves are excited. As discussed earlier, since the frequency of the driven mode satisfies $\omega_{\rm dr} > \omega_c $, it undergoes collisionless spatial evanescence: the intensity of the mode spatially decays as it propagates away from the wave source. The predicted spatial damping rate $\gamma_z \approx 0.4 k_p$ (indicated by a pink star atop of a blue line in Fig \ref{fig:Fig2}(d)) obtained from our semi-analytic model shows excellent agreement with the spatial damping observed in the simulation  (fitted black dot-dash line in Fig.~\ref{fig:Fig22}(d)). Thus, the surface plasma waves with frequencies in the $\omega_c < \omega < \omega_{+}$ spectral range are confined in all spatial dimensions: radially (at the plasma-vacuum interface) and axially (near their source) because of their spatial evanescence owing to their coupling to the continuum of upper-hybrid resonances localized inside the SPI.

\section{Propagation of TSPWs in axially-varying magnetic field} \label{section:expandingB}
In the preceding sections, we have assumed that the magnetized plasma is translationally invariant in the axial $z$-direction. Under such assumption, the external magnetic field is assumed to be uniform and unidirectional, and the axial wavenumber (either real-valued $k_z$ or complex-valued $\tilde{k}_z$) can be assumed to be conserved. Under such assumptions, a dispersion function $D_{\ast}\left( \tilde{\omega}, \tilde{k}_z, m \right)$ can be derived and various forms of the dispersion relations (e.g., $\tilde{\omega}(k_z,m)$ as in Fig.~\ref{fig:Fig2}(c) or $\tilde{k}_z(\omega,m)$ as in Fig.~\ref{fig:Fig2}(d)) can be derived from $D_{\ast}=0$. In this section, we explore spatio-temporal evolution of TSPW's in an expanding magnetic field. The presence of an inhomogeneous magnetic field introduces several fundamental questions that are absent in systems with a uniform external field. In particular, the propagation of TSPWs under spatially varying magnetic field conditions, i.e. when the translational symmetry in the axial direction is broken, has not been studied.

\begin{figure}[ht]
\centering
\includegraphics[width=1\textwidth]{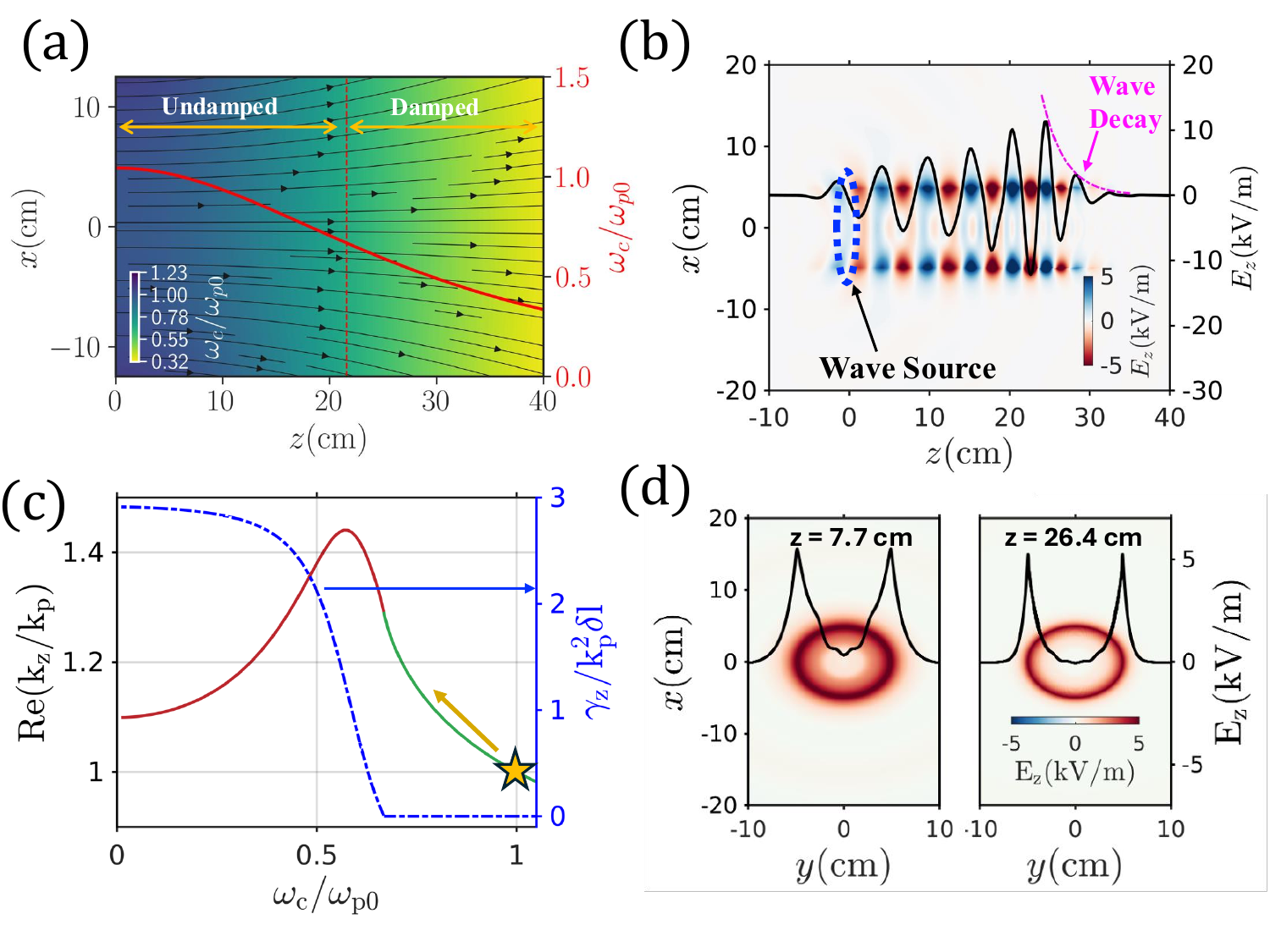}
\caption{\label{fig:Fig3} Propagation of TSPWs in axially non-uniform magnetic field. (a) The magnetic field distribution\cite{Simpson2001CircularLoop} generated in simulation box.  The solid red line represents the cyclotron frequency ($\omega_{c}(z)/ \omega_{p0}$) at $\rm  R_c = 4.8 cm$ ($k_p R_c = 5$).  The vertical red dashed line at $z = z_{\rm crit}\approx 21.6 \rm cm$ corresponds to \(  \omega_c(z_{\rm crit}) = \omega_{\rm dr}   =  2\pi \times 3.34 {\rm rad/s}\).  (b) A time snapshot of $\rm z-$directional electric field in $xz$-plane. The black solid line shows $E_z$ at $\rm  R = R_C = 4.8 {\rm cm}$. The TSPW undergoes damping around the resonance point $z_{\rm crit} $. (c) Complex-valued propagation wavenumber $\tilde{k}_z(\omega_{\rm dr},m=0)$ as a function of the cyclotron frequency $\omega_c$ for the driven mode $(\omega_{\rm  dr} = 0.67 \omega_{p0}$: undamped-TSPW (green solid line; left scale), damped TSPW (red solid line; left scale) and spatial damping rate (blue dot-dash line; right scale). The yellow star/line track the local $k_z$ for decreasing magnetic field. Parameters: bulk plasma density \( \rm n_{0} = 3 \times 10^{11}~\mathrm{cm^{-3}}\),  $\rm \omega_{p0}/2\pi = 5 GHz$,  \( \rm \lambda_p \approx 6~\mathrm{cm}\),  $\rm  R_{\rm c} = 4.8 cm$ ($k_{\rm p}  R_{\rm c}  = 5$), \(\delta l \approx 0.2 ~\mathrm{cm} (\rm k_p \delta_l = 0.2)  \),  $\rm k_z \approx 2 \pi /\lambda_p (k_{\rm z}  = k_{\rm p} )$(d) Representative snapshots of the electric field $E_z$ in the transverse $\rm xy$-plane at $\rm z = 7.7 cm$ ($\rm 26.4 {\rm cm}$) before (after) the resonant layer indicate stronger radial localization of the collisionlessly damped TSPW.}
\end{figure}

To investigate the effect of broken translational symmetry by non-uniform magnetic field, we use three-dimensional particle-in-cell (PIC) simulation code  \textsc{WarpX} ~\cite{WarpX}. The simulation box is $\mathrm{L_x = L_y \times L_z} =  \mathrm{20 ~\mathrm{cm} \times 20 ~\mathrm{cm} \times 90 \, ~\mathrm{cm}}$,  and has perfectly matched layers (PML's) on all six domain boundaries.  The box extends from $\rm z = -45 cm$ to $\rm z = 45 cm$. The external magnetic field is modeled as stationary but spatially inhomogeneous considering that a thin current carrying circular wire loop is kept at $\rm z = 0$, and for visualization plotted through its associated electron cyclotron frequency distribution in Fig.~\ref{fig:Fig3}(a). The magnetic field expressions produced by a thin current-carrying loop can be expressed as \cite{Simpson2001CircularLoop}
\begin{eqnarray} \label{eq:B_bottle}
  B_{0z} (r,z)  &= \frac{\mu_0 I_{\rm 0}}{\pi }  \frac{1}{2 \alpha^2 \beta}  \left(  \left(R_{\rm 0}^2 - r^2 -z^2 \right) E(\mathrm{k^2}) + \alpha^2 K(\mathrm{k^2}) \right), \nonumber   \\
 B_{0r} (r,z) &= \frac{\mu_0 I_{\rm 0}}{\pi }  \frac{z}{2 \alpha^2 \beta r}  \left( \left(R_{\rm 0}^2 + r^2 +z^2 \right) E(\mathrm{k^2}) - \alpha^2 K(\mathrm{k^2}) \right),
\end{eqnarray}
where $\alpha^2 \equiv R_{\rm 0}^2 + r^2 + z^2  - 2 r R_{\rm 0} $, $\beta^2 \equiv R_{\rm 0}^2 + r^2 + z^2  + 2 r R_{\rm 0} $, $k^2 \equiv 1 - \alpha^2/\beta^2  $,  $K$ and $E$ are complete elliptic integrals of first and second kind, $\mu_0$ is the vacuum permeability, $R_{\rm 0} = 40 {\rm cm}$ is the loop radius and $I_{\rm 0}$ is the current flowing in the loop.  The $I_{\rm 0}$ is chosen such that at \(R_c = 4.8~\mathrm{cm}\);\(z = 0 \) the electron cyclotron frequency is equal to bulk plasma frequency \( (\omega_c / \omega_{p0} = 1)\).

A Gaussian wave source with the field profile \[E_z \left( r,z,t \right) \propto \exp \left(- \left( \frac{r-d_{pos}}{\sigma_r} \right)^2 - \left( \frac{z} {\sigma_z} \right)^2 \right) \cos(k_z z -\omega_{\rm dr} t)\] is used to excite surface wave modes, where $d_{pos} = L_{int}$, $\rm \sigma_r = 5 \rm mm$, and $\rm \sigma_z = 2 \rm cm$.  As demonstrated in Section~\ref{section:theoretical model} (see Fig.~\ref{fig:Fig22}(b)), an axially thin source (\(\sigma_z \ll k_z^{-1}\)) is non-directional, i.e. it inevitably excites forward- and backward-propagating (\(\pm k_z\))  waves of equal magnitudes. To observe potential reflections of TSPW's propagating through non-uniformly magnetized plasma, it is necessary to minimize the excitation of a backward-propagating TSPW by the wave source itself. Therefore, the $\sigma_z$ is chosen in such a way that only the forward-moving TSPW couples to the plasma.

The results of the PIC simulation for the driven mode \( (\omega, m)= (0.67 \omega_{\rm p0}, 0) = (2\pi \times 3.34~\mathrm{rad/s}, 0)\) are shown in Fig.~\ref{fig:Fig3}(b), where only $k_{\rm forw}$ azimuthally symmetric mode ($m = 0$) couples to the plasma and propagates in the forward axial direction. As it propagates into a decreasing magnetic field, an increase in its axial wavenumber $k_z$ is indicated by the black line in Fig. \ref{fig:Fig3}(b)). This behavior agrees well with the isofrequency dispersion relation $\tilde{k}_z(\omega_c)$ (see Fig. \ref{fig:Fig3}(c)) obtained using our semi-analytic theory for different axial cyclotron frequencies and for a fixed $\omega = 0.67 \omega_{p0} = 2 \pi \times 3.34 \mathrm{rad/s}$. The remaining parameters are the same as in Fig. \ref{fig:Fig3} (indicated by the yellow star). Such increase in $k_z$ is accompanied by the decrease of the group velocity and increase of the wave amplitude, as shown by the black line in Fig.~\ref{fig:Fig3}(b).

Upon crossing the resonant layer near $\omega_c(z) \lesssim \omega_{\rm  dr}$ at $z_{\rm crit} \approx 21.6\,\mathrm{cm}$ (marked by the vertical red dashed line in Fig.~\ref{fig:Fig3}(a)), the wave undergoes collisionless damping and decays evanescently as observed in Fig.~\ref{fig:Fig3}(b) (labelled ''wave-decay").  The local dependence of the complex-valued propagation wavenumber $\tilde{k}_z$ in an axially-varying magnetic field is plotted in Fig.\ref{fig:Fig3}(c) as a function of the local cyclotron frequency $\omega_c$. The real part ($k_z$) of $\tilde{k}_z$ plotted as a red solid line indicates the compression of the wave period $\lambda_z \equiv 2\pi/k_z(\omega_c)$, followed by its rarefaction as the TSPW propagates deeper into the evanescence region corresponding to lower magnetic field. The corresponding spatial damping rate $\gamma_z(\omega_c)$ is plotted by dot-dash blue line in the same figure (right axis).

While the TSPW retains its azimuthal ($m=0$) symmetry in the evanescence region, its radial profile changes considerably as can be seen from Fig.~\ref{fig:Fig3}(d), where the spatial profiles of the axial electric field $E_z$ are plotted in the propagation (left figure) and evanescence (right figure) regions. The electric field enhancement of the collisionlessly-damped TSPWs around the resonance layer has been predicted for infinitely extended plasma slabs \cite{Rajawat_prl_2025}. As clearly indicated in Fig.~\ref{fig:Fig3}(d), this property is retained for finite-radius plasma columns immersed in non-uniform magnetic field: the damped TSPW in the evanescent region ($z=26.4 {\rm cm}$) acquires a more localized electric field profile than the original (undamped) TSPW in the propagation region ($z=7.7 {\rm cm}$).

Unlike the coupling of TSPW's with a single-resonance layer \( (r_{\rm UH} )\) discussed in the previous sections, here the wave encounters multiple resonant layers \( (r_{\rm UH}(r,z)/\delta l  = ( \omega_{\rm UH}^2 (r,z) )  - \omega)/\omega_{\rm p0}^2) \) as it propagates in the axial $z$-direction through the region of non-uniform magnetic field. This is because the local resonant upper-hybrid frequency \( \omega_{\rm UH}^2 (r,z) = \omega_c^2(z) + \omega_{\rm p}^2(\rm r) \) is the function of both radial and axial coordinates inside the SPI.  As demonstrated in Sec.~\ref{section:theoretical model}, our semi-analytic model accurately predicts the damping rate for a relatively smooth plasma–vacuum interfaces $\delta \ll k_p^{-1}$. Nevertheless in the presence of a continuum of resonances, the evanescent behavior of the TSPW cannot be described by a single spatial damping rate, as $\gamma_z$ becomes a function of $z$. Instead, a Budden tunneling coefficient $T(z) \sim \int_{z_{\rm crit}}^{z} dz^{\prime} \gamma_z(z^{\prime})$ more accurately quantifies the spatial decay of the mode \cite{Budden_1985}. Here $z_{\rm crit}$ is the critical position along the plasma satisfying $\omega_c(z_{\rm crit})=\omega_{\rm dr}$ and marked with a dashed line in Fig.~\ref{fig:Fig3}(a). In a decreasing magnetic field, the $m=0$ TSPW becomes evanescent at $z=z_{\rm crit}$ and continues its spatial decay for $z > z_{\rm crit}$.

It is worth noting that our PIC simulations properly account for both axial and radial variations of the axial magnetic field $B_{0z}(r,z)$ without making the simplifying narrow-plasma assumption. The presence of the finite radial component of the magnetic field given by Eq.(\ref{eq:B_bottle}) is also included. On the other hand, the simplified theory upon which the calculation of the complex-valued propagation wavenumber $\tilde{k}_z(\omega_c)$ plotted in Fig.~\ref{fig:Fig3}(c) is based takes neither of these into account. Nevertheless, the spatial patterns of the propagation and decay of the TSPW modeled by our PIC simulation and plotted in Fig.~\ref{fig:Fig3}(b) are in good agreement with those obtained by applying the simplified approach  of using the locally calculated $\gamma_z(z)$ to calculate the Budden tunneling coefficient. This suggests that the impact of the small but finite radial component $B_{0r}$ of the external magnetic field on the propagation of TSPWs down the gradient of the magnetic field is small. 

We note that the absence of any visible reflection of the forward-moving TSPW ($k_{\rm forw} = k_z > 0$) from the region of decreasing magnetic field persists even when the characteristic scale $L_z$ of the magnetic field decrease -- approximately equal to $L_z \sim R_0$ for the magnetic field of a current loop given by Eq.(\ref{eq:B_bottle}) -- satisfies $L_z \sim k_z^{-1}$. The lack of coupling between the forward and backward waves is due to the fact that the oscillatory part $k_z$ of the complex-valued propagation wavenumber $\tilde{k}_z$ remains finite even inside the evanescence region. That is in contrast to the waves propagating through over-dense unmagnetized plasma, where the spatial behavior of the wave is purely evanescent (i.e. non-oscillatory). For those cases, significant reflection is observed even when the plasma density is gradually increasing from under- to over-dense. In contrast, for the magnetic field configuration considered in this work (see Fig. \ref{fig:Fig3}(b)), gradual variation of the axial magnetic field guarantees that no reflected TSPWs with $k_{\rm back} = -k_z$ emerge. However, in regimes where the magnetic field exhibits an extremely sharp variation, i.e. $L_z < \left( 2k_z \right)^{-1}$, the broken axial symmetry becomes apparent, and the reflected waves are clearly seen in our PIC simulations (not shown).

\section{Conclusions}  \label{section:conclusions}
In summary, we have extended the theory of propagation and collisionless damping of topologically-protected surface plasma waves in magnetized plasmas from a simplified semi-infinite plasma slab geometry to the more realistic model of cylindrical plasmas in a non-uniform magnetic field. The temporal and spatial damping rates of the TSPWs spiraling around a magnetized plasma cylinder predicted by our semi-analytic model shows excellent agreement with the corresponding PIC simulations. We have also expanded our theory and numerical modeling from the commonly studied magnetized plasmas possessing exact translational symmetry to the regime of inhomogeneous three-dimensional magnetic fields produced by a realistic finite-sized current coils.  This enabled us to study an intriguing regime of TSPWs propagating downstream from the coils into the region of decreasing magnetic field amplitude while undergoing reflections-free collisionless damping. Despite the non-uniformity of such plasmas in the axial direction of the predominant magnetic field, it was found that such surface waves can have negligible reflection in the case of a gradual spatial change of the external magnetic field. The expanding magnetic field lines downstream from the coil produces an energy sink for the launched TSPWs localized in all three dimensions, into which they deposit their entire energy. In addition to heating the plasma via upper-hybrid heating \cite{Lin_pof_1982}, the absorbed TSPWs also impart a finite angular momentum to the plasma column\cite{Fu_jpp_2022}. In our future work, we will explore the possibility of imparting rotation to plasmas via angular momentum transfer by going beyond the current assumption of mobile electrons in the background of immobile ions. Exploring the new physics stemming from the finite ion mass \cite{ Rajawat_arxiv_2025} could be beneficial for a variety of applications requiring rotating plasmas, including isotope separation and improved plasma confinement.~\cite{rax_pop_2017}.

\ack{This work is supported by Air Force of Scientific Research (AFOSR) through Stanford University under MURI Award no. FA9550-21-1-0244.  The authors acknowledge the Texas Advanced Computing Center (TACC) at The University of Texas at Austin for providing computational resources that have contributed to the research results reported within this paper. The authors would like to thank Dr. Y. Raitses, PPPL for useful discussions.}

\section*{Appendix} \label{appendix}
We demonstrate collisionless damping of TSPW's in $(\theta,z)$-plane using 3D PIC simulations\cite{WarpX}. The simulation box is $\mathrm{L_x = L_y \times L_z} =  \mathrm{20 ~\mathrm{cm} \times 20 ~\mathrm{cm} \times 60 \, ~\mathrm{cm}}$, and has perfectly matched layers (PML's) on all six domain boundaries. The box extends from $\rm z = -15 cm$ to $\rm z = 45  cm$. A point Gaussian wave source as illustrated in Fig. \ref{fig:Fig1}(a) and it's mathematical form along with temporal envelope parameters envelope are identical to those used in the section \ref{section:expandingB}. The point-source frequency is $\rm \omega_{\rm dr} = 0.67 \omega_{p0} = 2 \pi \times 3.34 \, rad/sec$. 

\begin{figure}[h!]
    \centering
    \includegraphics[width=0.8\textwidth]{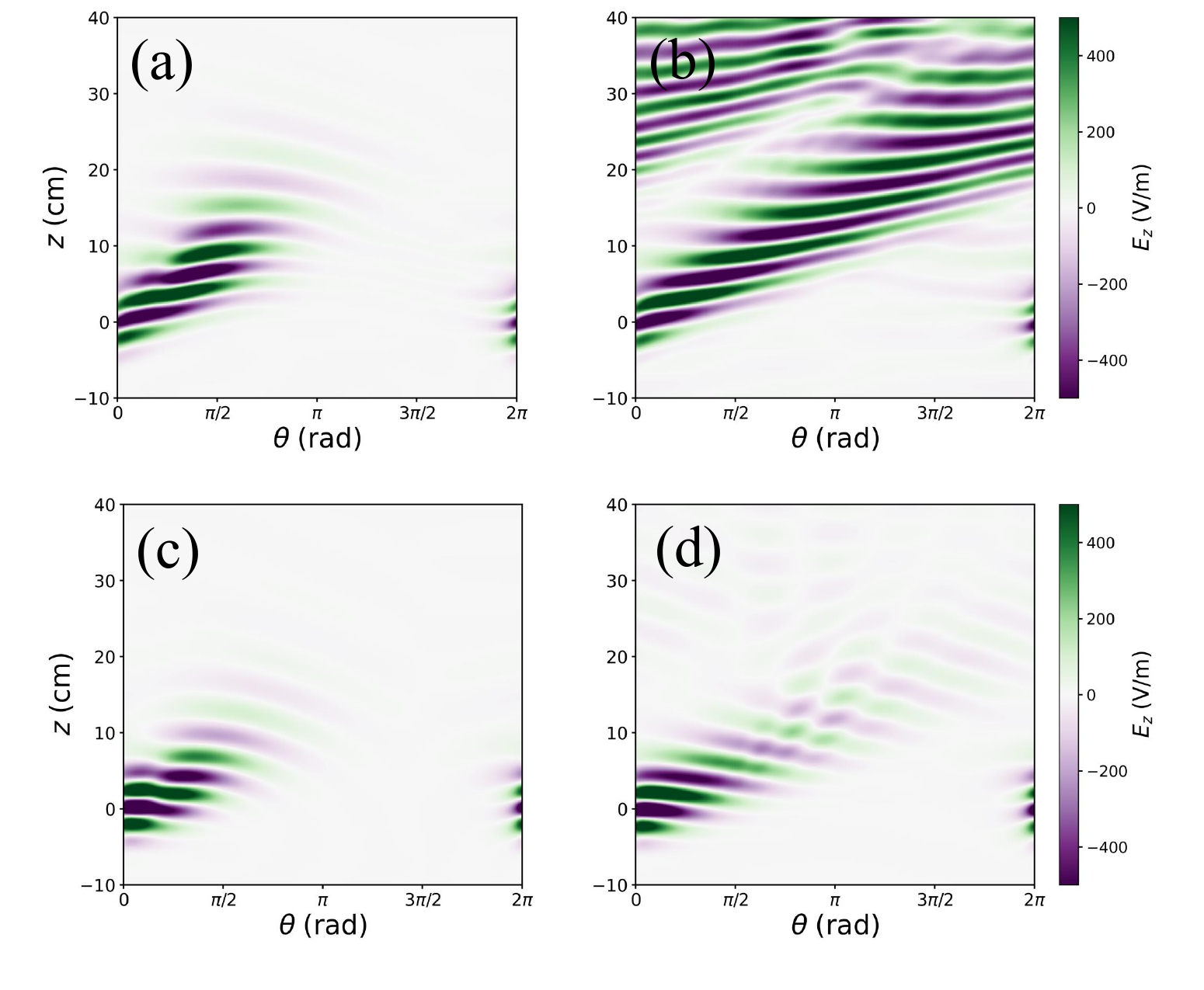}
    \caption{ Time snapshots of $\rm z-$directional electric field ($E_z$) in $(\theta,z)$-plane for (a,b) $\omega_c = \omega_{p0}$ and (c,d) $\omega_c = 0.5 \omega_{p0}$. Time snapshots are plotted at (a,c) $T = 2 \, T_{\rm TSPW}$ and (b,d) $T = 7.5 \, T_{\rm TSPW}$, where $\rm T_{\rm TSPW} \approx 1.25 ns$. The field values are interpolated on $(\theta,z)$ surface from $\rm r \approx 5.1 cm$. Common plasma parameters: same as in Fig.~\ref{fig:Fig3}.}
    \label{fig:Fig4}
\end{figure}

We consider two cases based on different cyclotron-frequencies. For $\omega_c = \omega_{p0}$, the point-source–excited TSPWs initially spiral around the plasma column (Fig. ~\ref{fig:Fig4}a) and, at later times, relax toward an $m \simeq 0$ mode (Fig.~\ref{fig:Fig4}b). In contrast, for $\omega_c = 0.5\,\omega_{p0}$ the TSPWs again begin to spiral (Fig.~\ref{fig:Fig4}c), but since $\omega_{\rm dr} > \omega_c$ they become evanescent due to collisionless damping and do not develop an $m \simeq 0$ mode (Fig.~\ref{fig:Fig4}d).
 
\bibliography{cylindrical_topo_spw.bib}

\end{document}